\documentclass{aa} 

\usepackage{txfonts}
\usepackage{natbib}
\usepackage{graphicx}
\usepackage{color}
\usepackage{soul}

\definecolor{red}{rgb}{1,0,0}
\definecolor{blue}{rgb}{0,0,1}
\definecolor{green}{rgb}{0,1,0}
\newcommand{\rouge}{}

\bibpunct{(}{)}{;}{a}{}{,}

\title{FIRST, a fibered aperture masking instrument}
\subtitle{II. Spectroscopy of the Capella binary system at the diffraction limit}
\titlerunning{Spectroscopy of the Capella binary system at the diffraction limit}

\author{
   E.~Huby\inst{1}
\and G.~Duch\^ene \inst{2,3}
\and F.~Marchis \inst{4}
\and S.~Lacour \inst{1}
\and G.~Perrin \inst{1}
\and T.~Kotani \inst{5}
\and \'E.~Choquet \inst{1,6}
\and E.~L.~Gates \inst{7}
\and O.~Lai \inst{8,9}
\and F.~Allard \inst{10}
}
\institute{
     LESIA, Observatoire de Paris, CNRS, UPMC, Universit\'e Paris-Diderot, Paris Sciences et Lettres, 5 place Jules Janssen, 92195 Meudon, France
\and Department of Astronomy, University of California at Berkeley, Hearst Field Annex, B-20, Berkeley CA 94720-3411, USA
\and UJF-Grenoble 1 / CNRS-INSU, Institut de Plan\'etologie et d'Astrophysique (IPAG) UMR 5274, 38041 Grenoble, France
\and Carl Sagan Center at the SETI Institute, 189 Bernardo Av., Mountain View CA 94043, USA
\and Extrasolar Planet Detection Project Office, National Astronomical Observatory of Japan, 2-21-1 Osawa, Mitaka, Tokyo 181-8588, Japan
\and Space Telescope Science Institute, 3700 San Martin Drive, Baltimore, MD 21218, USA
\and University of California Observatories/Lick Observatory, P.O. Box 85, Mount Hamilton, CA 95140, USA
\and Gemini Observatory, 670 North A'ohoku Place Hilo, Hawaii 96720, USA
\and Subaru Telescope, 650 North A'ohoku Place, Hilo, Hawaii 96720, USA
\and CRAL UMR 5574: CNRS, Universit\'e de Lyon, \'Ecole Normale Sup\'erieure de Lyon, 46 all\'ee d'Italie, 69364 Lyon Cedex 7, France
}

\abstract
{
}
{
FIRST is a prototype instrument built to demonstrate the capabilities of the pupil remapping technique, using single-mode fibers and working at visible wavelengths. Our immediate objective is to demonstrate the high angular resolution capability of the instrument and to show that the spectral resolution of the instrument enables characterisation of stellar companions.
}
{
The FIRST-18 instrument is an improved version of FIRST-9, that simultaneously recombines two sets of nine fibers instead of one, thus greatly enhancing the ($u$,$v$) plane coverage. We report on observations of the binary system Capella at three epochs over a period of 14 months ($\gtrsim$4 orbital periods) with FIRST-18 mounted on the 3-m Shane telescope at Lick Observatory. The binary separation during our observations ranges from 0.8 to 1.2 times the diffraction limit of the telescope at the central wavelength of the spectral band.
}
{
We successfully resolved the Capella binary system at all epochs, with an astrometric \rouge{precision} 
as good as 1\,mas under the best observing conditions. FIRST also gives access to the spectral flux ratio between the two components directly measured with an unprecedented spectral resolution of $R\sim300$ over the 600--850\,nm range. In particular, our data allow to detect the well-known overall slope of the flux ratio spectrum, leading to an estimation of the "pivot" wavelength of 0.64$\pm$0.01$\mu$m, at which the cooler component becomes the brightest. Spectral features arising from the difference in effective temperature of the two components (specifically the H$\alpha$ line, TiO and CN bands) have been used to constrain the stellar parameters. The effective temperatures we derive for both components are slightly lower (5--7\%) than the well-established properties for this system. This difference mainly originates from deeper molecular features than those predicted by state-of-the-art stellar atmospheric models, suggesting that molecular line lists used in the photospheric models are incomplete and/or oscillator strengths are underestimated, most likely concerning the CN molecule.
}
{
These results demonstrate the power of FIRST, a fibered pupil remapping based instrument, in terms of high angular resolution and show that the direct measurement of the spectral flux ratio provides valuable information to characterize little known companions.
}


\begin{document}

\maketitle

\section{Introduction}

The objective of the FIRST (Fibered Imager foR a Single Telescope) prototype development is the detection of faint companions such as exoplanets. Many instruments currently being designed and manufactured are addressing this challenge, which requires high dynamic ranges at high angular resolution. Different solutions are implemented such as extreme adaptive optics systems, coronagraphic masks and interferometric techniques (or a combination thereof). FIRST belongs to the latter category since its principle relies on the aperture masking \citep{Haniff1987}, in which a mask with small holes is put on the pupil of the telescope. In a traditional implementation, these holes are organized in a non redundant manner in order to avoid the fringe blurring due to the atmospheric turbulence, which leads to the loss of most of the high spatial frequency information on the object. The image is thus the superimposition of all fringe patterns coded with as many spatial frequencies. Object visibilities corresponding to every sub-pupil pairs can therefore be retrieved. This technique has been established as a standard for diffraction limited observations, up to dynamic ranges of a few hundreds, thanks to the routinely used Sparse Aperture Masking mode of the NACO instrument at the VLT \citep{Lagrange2012, Sanchez-Bermudez2012, Grady2013, Cieza2013} and \rouge{the} Keck NIRC2 aperture masking experiment \citep{Hinkley2011, Blasius2012, Evans2012}.

FIRST is the implementation of the fibered pupil remapping technique as proposed by \cite{Chang1998} and then \cite{Perrin2006} at visible wavelengths. This technique can be seen as a variant of the aperture masking technique
that aims at increasing the achievable dynamic range. As stated by \cite{Baldwin2002}, dynamic range directly depends on the number of sub-pupils and on the accuracy of the observable measurements (visibilities or closure phases for instance). In the FIRST instrument, single-mode fibers offer a way to improve these two aspects: (i) they allow to use the whole telescope pupil while their outputs can be non-redundantly recombined and (ii) they spatially filter the wavefront \rouge{and} thus avoid 
speckle noise. The spatial phase fluctuations due to the atmospheric turbulence are thus traded against flux fluctuations because of the imperfect coupling efficiency into the fibers. However, these can be more easily handled during data reduction than phase fluctuations. A self-calibration algorithm has indeed been proposed to retrieve the complex visibilities without the need of specific photometric channels \citep{Lacour2007}.

The first light of the instrument was achieved in July 2010 at Lick Observatory with FIRST-9 mounted on the Cassegrain focus of the 3-m Shane telescope \citep{Huby2012}. The instrument has since been significantly improved to increase the number of sub-pupils and the accuracy on the closure phase measurements, and hence the achievable dynamic range. FIRST-9 has thus become FIRST-18, in which 18 sub-pupils are recombined, and has been mechanically enhanced to reach a higher stability. Four observing runs have been conducted between October 2011 and December 2012. Several binary systems have been observed, as these simple objects are particularly suited for testing the capabilities of the instrument.

Among these binaries, Capella ($\alpha$ Aurigae, $V$=0.08, $R$=-0.52) has been observed at three different epochs with FIRST-18. Capella is a well-known nearby binary system, that has been studied for more than a hundred years, and its historical record is \rouge{notably} 
linked to the Lick Observatory and the first measurements with interferometers. The first announcement of discovery of this spectroscopic binary dates indeed from \cite{Campbell1899} who used the Mills spectrograph installed on the 36-inch Lick refractor to observe Capella (simultaneously \cite{Newall1899} made the same discovery using the Cambridge spectroscope). Capella later was the first binary system \rouge{whose} 
astrometric orbit has been interferometrically measured with the 6-m baseline Michelson interferometer on the 100-inch telescope at Mount Wilson \citep{Anderson1920, Merrill1922}. For decades, this friendly target has been observed with various interferometers and speckle imaging techniques. So far, the most accurate measurements of the astrometric orbit have been performed by \cite{Hummel1994} using the Mark III interferometer at Mount Wilson, using baselines from 3\,m up to 23.6\,m. Concerning the abundant bibliography about Capella, \cite{Torres2009} provide a very complete review of all spectroscopic, as well as interferometric measurements available at that time and they use them to derive the parameters of the system: effective temperatures of 4920 $\pm$ 70\,K and 5680 $\pm$ 70\,K, radii of 11.87 $\pm$ 0.56 $R_{\odot}$ and 8.75 $\pm$ 0.32 $R_{\odot}$, luminosities of 79.5 $\pm$ 4.8 $L_{\odot}$ and 72.1 $\pm$ 3.6 $L_{\odot}$ and the parameters of the relative orbit. The latest and most accurate masses have been determined by \cite{Weber2011} using the spectroscopic orbit combined with the inclination of the orbital plane derived from astrometry measurements: 2.573~$\pm$~0.009~$M_{\odot}$ and 2.488~$\pm$~0.008~$M_{\odot}$.

With an angular separation varying from 41\,mas to 56\,mas, Capella is therefore an ideal target to probe the capabilities of FIRST-18 at the diffraction limit, which ranges from 41 mas at 600\,nm to 58\,mas at 850\,nm for a 3-m telescope. In this paper, we report on the successful detection of Capella as a binary system at the diffraction limit of the telescope using FIRST-18, which is described in Sect. \ref{sec_Obs_data_red}, along with the data reduction procedure. The results are presented in Sect. \ref{sec_Results}. The spectrally-dispersed flux ratio measurement is of particular interest as it is the first time that it is {\sl directly} measured with a $R\sim300$ spectral resolution. Higher-resolution spectra of the two components have been estimated by "deblending" the composite spectrum of the binary using spectral templates to disentangle them \citep{Barlow1993}. Flux ratio measurements of the system, on the other hand, have been obtained in discrete broad- and narrow-band filters that are too sparsely distributed to provide fine spectral information \citep{Torres2009}. Thus, the FIRST data provide a unique insight \rouge{into} 
the Capella system and, by extension, all systems with separation as small as the diffraction limit of a given telescope. In this work, we model our FIRST flux ratio spectrum to constrain the effective temperatures of the Capella binary, thereby demonstrating the power of FIRST to study previously uncharacterized companions in the future.

\section{Observations and data reduction} \label{sec_Obs_data_red}

Observations were conducted with the 3-m Shane telescope \rouge{at} 
Lick Observatory from 2011 to 2012 (see Table \ref{tab_Obslog}). As in previous observations \citep{Huby2012}, the Shane adaptive optics system provided sufficient correction to stabilize the fringes, although it is optimized for the infrared wavelengths.

\subsection{FIRST-18}

\begin{figure}
\centering
\resizebox{\hsize}{!}{\includegraphics{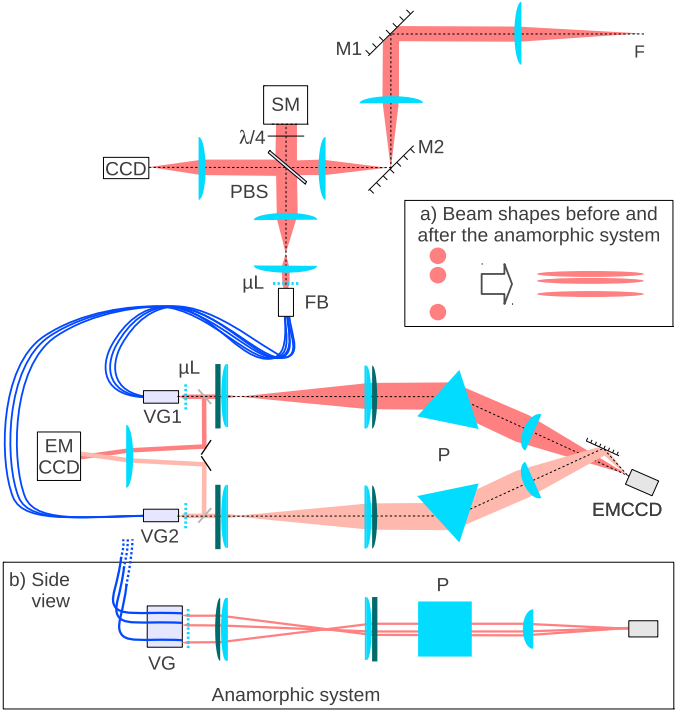}}
\caption{Set-up of the FIRST-18 instrument. The injection part of the set-up, up to the fiber bundle, is basically unchanged compared to the previous version of the instrument. 
The recombination part on the other hand has been duplicated. The EMCCD dedicated to the measurement of the transmitted flux during the optimization procedure has been also depicted. The mirrors drawn as gray lines (just after the V-grooves and microlens arrays) are removable from the light path. Inset a) illustrates the resulting beam shapes after the anamorphic system. Inset b) is a side view of the recombination of the beams (only three beams out of nine are represented). SM: Segmented mirror. PBS : Polarizing beam splitter. $\mu L$: Microlens array. FB: Fiber bundle. VG: V-groove. P: Dispersing prism.} \label{fig_setup}
\end{figure}

\begin{figure}
\centering
\resizebox{\hsize}{!}{\includegraphics{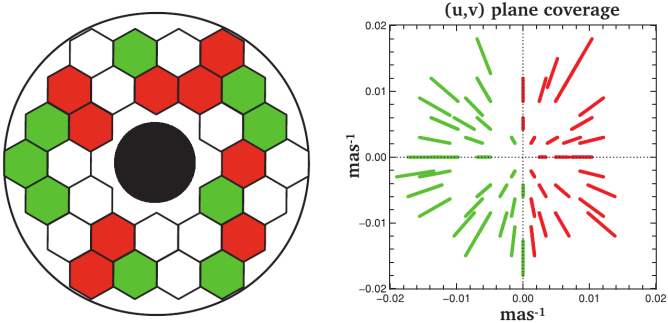}}
\caption{Fiber configurations of the entrance pupil, the green and red colors corresponding to each of the two sets of nine fibers. The corresponding polychromatic ($u$,$v$) plane coverage is represented on the right panel. For \rouge{clarity,} 
the ($u$,$v$) plane coverage of each 9-fiber configuration is not symmetrized.
}
\label{fig_config}
\end{figure}

After the first light of the instrument obtained in 2010 \citep{Huby2012}, efforts have been focused on improvements to the stability, sensitivity and dynamic range achievable with FIRST. The injection part of the instrument (including the pupil imager, the Iris AO segmented mirror used to precisely steer the beams on the fiber cores that are gathered in the fiber bundle, from Fiberguide Industries) is basically unchanged. However significant upgrades have been carried out in the recombination part of the instrument, as can be seen in Fig. \ref{fig_setup}.

Assuming that the noise is uncorrelated between the baselines -- a reasonable assumption in the photon noise limited regime -- the dynamic range increases with the number of visibility and closure phase measurements \citep{Baldwin2002,Lacour2011,Lebouquin2012}. The main development was therefore aimed at increasing the number of sub-pupils, leading to the FIRST-18 instrument, in reference to the two sets of 9 sub-pupils that are recombined. The non-redundancy of the linear recombination configuration is a technical limit: standard V-groove chips offer 48 spaces where fibers can be positioned. The most compact 9-fiber configuration requires $n_\mathrm{max}=$44 spaces, while the 10-fiber and 11-fiber ones require respectively 55 and 75 spaces. In addition, the 9-fiber configuration is also a compromise between the available space on the bench (defining the focal length $f^{\prime}$ of the imaging lens), the diffraction pattern width ($f^{\prime} \lambda/D$, with $D$ corresponding to the V-groove pitch, and also to the collimated beam size at the fiber output) and the sampling of the highest frequency fringe pattern (of period $T=f^{\prime}\lambda/\left(n_\mathrm{max} D\right)$) that fits the detector dimensions. For these reasons, two sets of 9 fibers are recombined independently. Given the 30 available sub-pupils in the obstructed telescope pupil (see Fig. \ref{fig_config}), our selected fiber configuration using 18 of them
gives access to 73\% of all possible independent baselines. As the instrument is mounted at the Cassegrain focus of the telescope, the ($u$,$v$) plane coverage does not rotate during the night and is stable in time. However, the ($u$,$v$) plane coverage is significantly extended thanks to the broad wavelength range, as illustrated in Fig. \ref{fig_config}.

Each recombination scheme of the beams is built on the same design as for the previous version. The collimated beams coming from the V-grooves are reshaped thanks to anamorphic systems that mimic the slit of a spectrometer consisting in dispersing prisms (see Fig. \ref{fig_setup}). The anamorphic system has been designed to satisfy the Nyquist-Shannon sampling condition for the highest frequency fringe pattern (for a given focal length of the imaging lens) on the one hand, and to reach a high spectral resolution on the other. This is achieved thanks to a 4-lens (2 spherical and 2 cylindrical lenses) afocal system performing a beam elongation of 20 in the direction of dispersion and a compression of 7 in the orthogonal direction. The two fringe patterns are imaged on the same detector side by side along the wavelength axis. Spectral filters have been inserted in each arm in order to avoid overlapping of the patterns. In the center of the detector, the fringe patterns are cut at 600\,nm and 850\,nm thereby defining the spectral range of the instrument.

Improvement of the data quality has also been provided by the use of a more efficient detector\rouge{, an Hamamatsu EMCCD camera, with maximal EM gain of 1200, quantum efficiency of 90\,\%, 77\,\% and 54\,\% at 600, 700 and 800\,nm, respectively, and}
%
512 $\times$ 512 pixels 16\,$\mu \mathrm{m}$ wide. The overall mechanical stability has been enhanced too, allowing to observe objects up to 40\degr\,from zenith without significant loss of flux due to mechanical flexures.

\subsection{Acquisition procedure}

\begin{table}
\centering
\caption{Observation log, including the dates of observation, the type of 
target (AC stands for astrometric calibrator and C for closure phase calibrator), the number of images acquired with \rouge{an} 
integration time $t_{int}$ and an estimate of the $r_0$ parameter (evaluated thanks to the AO system occasionally once or twice during one sequence of observations).
}
\begin{tabular}{ccccc}
\hline \hline
Target & Type & $N_{img}$ & $t_{int}$ (ms) & $r_0$ at 550\,nm\\
\hline
\multicolumn{5}{c}{2011-10-16} \\
\hline
Algol     & AC & 10,000 & 150 & $\sim$ 11.5\,cm\\
Aldebaran & C & 10,000 & 50 & $\sim$ 10\,cm\\
Capella   & & 16,000 & 50 & $\sim$ 11.5\,cm\\
\hline
\multicolumn{5}{c}{2011-10-17} \\
\hline
Aldebaran & C & 10,000 & 50 & $\sim$ 16\,cm\\
Capella   & & 11,000 & 50 & $\sim$ 15\,cm\\
\hline
\multicolumn{5}{c}{2011-10-19} \\
\hline
Algol     & AC & 10,000 & 200 & $\sim$\,14 cm\\
Aldebaran & C & 10,000 & 100 & $\sim$ 8\,cm\\
Capella   & & 12,000 & 100 & $\sim$ 13\,cm\\
\hline
\multicolumn{5}{c}{2012-07-29} \\
\hline
$\alpha$ Per & C & 5,000 & 150 & $\sim$ 14\,cm\\
Algol     & AC & 5,000 & 150 & $\sim$ 15\,cm\\
Capella   & & 5,000 & 150 & $\sim$ 12.5\,cm\\
\hline
\multicolumn{5}{c}{2012-12-19} \\
\hline
Algol     & AC & 6,000 & 200 & $\sim$ 7\,cm\\
$\alpha$ Per & C & 5,000 & 200 & $\sim$ 8\,cm\\
Capella   & & 5,000 & 200 & $\sim$ 5\,cm\\
$\alpha$ Per & C & 5,000 & 200 & $\sim$ 7\,cm\\
$\beta$ Aur  & C & 5,000 & 200 & $\sim$ 8\,cm\\
Capella   & & 5,000 & 200 & $\sim$ 8\,cm\\
$\beta$ Aur  &  C & 5,000 & 200 & $\sim$ 8\,cm\\
\hline
\multicolumn{5}{c}{2012-12-20} \\
\hline
$\beta$ And  & C & 4,000 & 200 & $\sim$ 8\,cm\\
Algol     & AC & 6,000 & 200 & $\sim$ 8\,cm\\
$\alpha$ Per & C & 5,000 & 200 & $\sim$ 6\,cm\\
Capella   & & 5,000 & 200 & $\sim$ 7\,cm\\
\hline
\end{tabular}
\label{tab_Obslog}
\end{table}

Capella was observed in October 2011, July 2012 and December 2012 with FIRST-18 mounted on the 3--m Shane telescope at Lick Observatory. The observation log is reported in Table \ref{tab_Obslog}. Several bright ($0 \lesssim R \lesssim 3$) single stars were observed immediately before/after Capella so serve as calibrator for closure phase measurements and baseline calibration (see Sect. \ref{seq_orbit}).

The acquisition procedure is similar to the description given in \cite{Huby2012}. One data set comprises fringe sequences, fiber calibration files (every fiber diffraction pattern is recorded individually), and background files. However, this procedure has been significantly shortened compared to the previous observations thanks to the implementation of a dedicated optimization path in the optical set-up. The injection of the flux into the fibers is indeed very sensitive and must be optimized at every target 
switch, to compensate for residual mechanical flexures occurring in the instrument but also in the AO system. While this optimization was done one fiber at a time in 2010 (by scanning the corresponding micro-segment and measuring the transmitted flux directly on the science camera), it is now done simultaneously for all fibers thanks to a motorized mirror system that allows to image all the output fibers on a dedicated detector. The 18 corresponding micro-segments can therefore be scanned at the same time. The optimization is now completed in less than two minutes leading to a substantial gain in time and stability. 

\subsection{Data reduction}
\label{seq_data_red}

\begin{figure}
\centering
\resizebox{7cm}{!}{\includegraphics{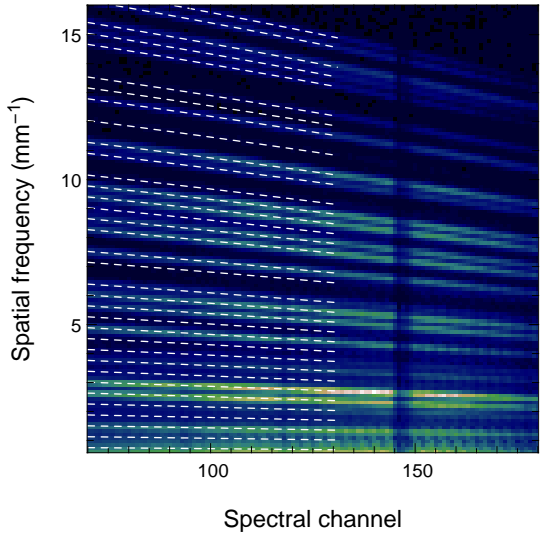}}
\caption{Mean power spectral density computed from 5000 50\,ms-images for Capella on 2011 Oct. 16 using data from a single V-groove. The color-scale has been adjusted discarding the zero-frequency peak (not visible on the image). The fitted peak positions are superposed as dashed white lines, only on the left part of the image for better visibility. The (telluric) absorption line appearing at spectral channel number 147 is exactly vertical, showing that the distortion in the image has been effectively corrected. At least 27 different peaks are clearly visible out of the 36 expected ones.
}
\label{fig_Capella_DSP}
\end{figure}

The basic principles of the FIRST data analysis are based on the P2VM method (\textit{Pixel to Visibility Matrix}), that was first developed for the AMBER instrument at the VLTI \citep{Millour2004} and subsequently adapted to FIRST data, as described in more detail in \cite{Huby2012}. The current data reduction process consists of the following steps:
\begin{itemize}
\item background (including dark current) subtraction ;
\item \rouge{optical} distortion correction ; 
\item spectral calibration based on the fitting of telluric absorption lines and features of the stellar spectrum ;
\item fitting of the fringe spatial frequencies by adjusting the peak positions in the mean power spectral density ;
\item calibration of the P2VM matrix from individual fiber profiles and spatial frequencies ;
\item pseudo-inversion of the P2VM matrix and computation of the best parameter sets in the least-squares sense ;
\item closure phase computation ;
\item closure phase calibration by measurements obtained on an unresolved target.
\end{itemize}

The distortion mentioned in the \rouge{second} 
step of the reduction results from astigmatism introduced by the prism, and has an amplitude of about 10-15 pixels. The shape of this distortion is evaluated by detecting the position of the deepest telluric absorption line and is corrected by horizontally interpolating and shifting the image values.

The spectral calibration is done using the UVES sky emission atlas \cite{Hanuschik2003}. As the light is spectrally dispersed through an SF2-prism, there are two parameters defining the wavelength as a function of the pixel: the angle of incidence on the prism $i$, which affects the spectral resolution, and the central wavelength $\lambda_0$. The continuum of the spectrum is fitted and subtracted, leaving only (stellar and telluric) absorption features. The same procedure is applied to the synthetic spectrum, and the parameters $i$ and $\lambda_0$ are retrieved by maximizing the correlation product between the two spectra. A fine adjustment is then performed by fitting the peak positions of the power spectral density computed over all images. The results of this fit is shown in Fig. \ref{fig_Capella_DSP} where the fitted peak positions are superposed to the mean power spectral density.

After applying the P2VM method, the complex coherent fluxes are thus retrieved and their phases are combined to estimate closure phases. The data reduction output comprises 84 closure phase measurements for each of \rouge{about} 
180 spectral channels from 600\,nm to 850\,nm. Since spatial frequency depends on wavelength, specific signal appears in the closure phase plotted as a function of wavelength, depending on the structure of the target.

\begin{figure*}
\centering
\includegraphics[width=18cm]{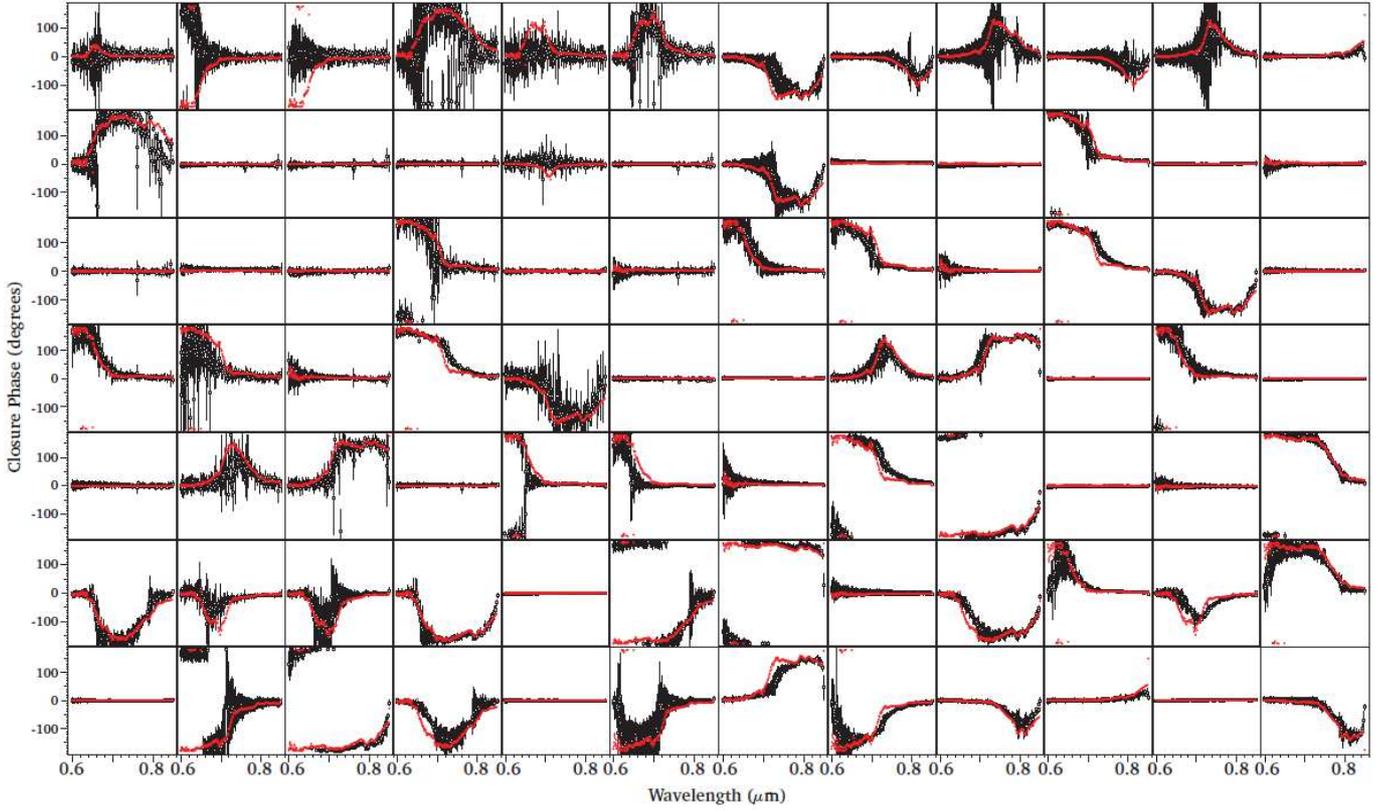}
\caption{Closure phase measurements obtained on 19$^{th}$ October 2011 (black points with rescaled error bars such that the reduced $\chi^2$ per spectral channel is unity) and best fit model (red points) computed from the best estimates of the binary parameters.
}
\label{fig_CP_Capella}
\end{figure*}

\subsection{Fitting of a binary model}

The final step in the analysis consists of fitting the closure phase estimates with a binary model. Three parameters are optimized by minimizing the $\chi^2$ function: two angular parameters defining the position of the companion, $\alpha$ and $\delta$, and a 
flux ratio \rouge{for each} 
spectral channel, $\rho(\lambda)$. The closure phase model function corresponding to a binary directly reads \citep{Lebouquin2012}:
\begin{equation}
f(\alpha, \delta, \rho) = \arg \left( \frac{(1+\rho e^{i \alpha_{ij}})(1+\rho e^{i \alpha_{jk}})(1+\rho e^{i \alpha_{ki}})}{(1+\rho)^3} \right),
\end{equation}
with
\begin{equation}
\alpha_{ij}(\alpha, \delta)=2 \pi \left( \alpha u_{ij} +  \delta v_{ij} \right),
\end{equation}
\noindent where $u_{ij}$ and $v_{ij}$ are the projections of the spatial frequency vector $\overrightarrow{s_{ij}}$ corresponding to the sub-pupils $i$ and $j$, on two orthonormal axes defined by the segmented mirror reference frame (see Sect.~\ref{seq_orbit} for details concerning the baseline calibration). This vector expresses as $\overrightarrow{s_{ij}}=\overrightarrow{B_{ij}}/\lambda$ with $\overrightarrow{B_{ij}}$ the baseline vector represented in the telescope pupil.

The $\chi^2$ function is therefore 3 dimensional. For every spectral channel, the contributions of all $n_\mathrm{CP}$ closure phases are added quadratically:
\begin{equation}
\chi^2_{\lambda}(\alpha, \delta, \rho) = 
\sum_k ^{n_\mathrm{CP}}
\frac{{\Delta CP_{\lambda}^k}^2} 
{{\sigma_{\lambda}^k}^2},
\end{equation}
with $\sigma_{\lambda}^k$ the error on closure phase $k$ at wavelength $\lambda$ and 
\begin{equation}
\Delta CP_{\lambda}^k= CP_{\lambda}^k - {f}_{\lambda}^k (\alpha, \delta, \rho) \pmod {2 \pi},
\label{eq_modpi}
\end{equation}
is the phase difference between the data and the model, defined modulo $2 \pi$.

Under the assumption that the noise affecting the data follows a Gaussian distribution, the likelihood function can be derived from the $\chi^2$ function by:
\begin{equation}
{\cal{L}}_{\lambda} (\alpha, \delta, \rho) \propto \exp\left(-\frac{\chi^2_{\lambda}(\alpha, \delta, \rho) }{2}\right).
\end{equation}

The proportional symbol is used as the likelihood function is normalized afterwards.

The optimal value for a given parameter can then be retrieved after marginalizing the likelihood function, i.e. by integrating over all other parameters. For the $\alpha$ parameter for instance, it is expressed as:
\begin{eqnarray} \nonumber
{\cal{L}}_\lambda (\alpha) = \iint {\cal{L}_\lambda} (\alpha, \delta, \rho) \mathrm{d} \delta \mathrm{d} \rho.
\end{eqnarray}

Similarly, a circular permutation over the set of parameters leads to ${\cal{L}}_\lambda (\delta)$ and ${\cal{L}}_\lambda (\rho)$. For every spectral channel, the optimal parameter sets $\left( \tilde{\alpha}_\lambda, \tilde{\delta}_\lambda, \tilde{\rho}(\lambda)\right)$ are then determined by the median value of these probability densities, and the associated error bars are defined as the 68\% confidence interval.

The $\lambda$ dependence has been noted differently for the position parameters and the flux ratio since the position of the companion is achromatic.
The $\tilde{\alpha}_\lambda$ and $\tilde{\delta}_\lambda$ have therefore been averaged in order to determine the final position $\left( \tilde{\alpha}, \tilde{\delta} \right)$. Around 20\,\% of the data points with the worst error bars have not been taken into account to compute the averaged position, in order to discard aberrant points. The associated error bars on $\tilde{\alpha}$ and $\tilde{\delta}$ are computed from the standard deviation of the respective distributions. 

\section{Results}
\label{sec_Results}

\begin{table*}
\centering
\caption{Companion position measurements for the outer Algol system (C relative to A+B) converted into angular separation and position angle. The ratio between the measured and theoretical angular separations give an estimate of the corrective scaling factor. The differences in position angle indicate the rotation of the segmented mirror reference frame relative to the North-East orientation. Horizontal lines separate the three observing runs considered here. Correction factors are assumed to be constant within a given observing run.}
\begin{tabular}{cc|ccc|ccc}
\hline \hline
&  & \multicolumn{3}{c}{$\rho$ (mas)} & \multicolumn{3}{c}{$\theta$ (\degr)}  \\
Date & JD (2,400,000+) & \rouge{Estimated} & \rouge{Predicted} & Ratio & \rouge{Estimated} & \rouge{Predicted} & Difference  \\
\hline
2011-10-16 & 55851.9 & 87.9  $\pm$ 9.8 & 78.0  $\pm$ 0.2 & 0.888 $\pm$ 0.099 
        & 218.1 $\pm$ 7.3 & 134.4 $\pm$ 0.1 & -83.7 $\pm$ 7.3 \\
2011-10-19 & 55854.9 & 77.6  $\pm$ 1.1 & 77.8  $\pm$ 0.2 & 1.002 $\pm$ 0.014 
        & 215.5 $\pm$ 0.7 & 134.7 $\pm$ 0.1 & -80.9 $\pm$ 0.7 \\
\hline
2012-07-29 & 56139.0 & 84.4 $\pm$ 1.4 & 88.0  $\pm$ 0.5 & 1.043 $\pm$ 0.019 
        & 13.4 $\pm$ 1.4 & 307.4 $\pm$ 0.1 & -66.0 $\pm$ 1.4 \\
\hline
2012-12-19 & 56281.7 & 89.0 $\pm$ 5.6 & 94.0  $\pm$ 0.4 & 1.056 $\pm$ 0.067 
        & 13.4 $\pm$ 4.1 & 314.5 $\pm$ 0.1 & -58.9 $\pm$ 4.1 \\
2012-12-20 & 56282.8 & 94.6 $\pm$ 6.7 & 93.6  $\pm$ 0.4 & 0.989 $\pm$ 0.070 
        & 15.5 $\pm$ 5.4 & 314.5 $\pm$ 0.1 & -60.9 $\pm$ 5.4 \\
\hline
\label{tab_Algol}
\end{tabular}
\end{table*}

\begin{table*}
\centering
\caption{Angular parameter estimates for Capella at different observation dates. The reference axes are defined by the segmented mirror orientation which is actually rotated compared to the North-East orientation. The uncorrected estimates do not take the rotation of the image plane and the scaling factor into account, while the corrected ones are the final estimates. Horizontal lines separate the three observing runs considered here.}
\begin{tabular}{cc|ccc|ccc}
\hline \hline
& & \multicolumn{3}{c}{Angular separation (mas)} &  \multicolumn{3}{c}{Position angle (\degr)}\\
Date & JD (2,400,000+) & Uncorrected & Corrected & Predicted & Uncorrected & Corrected & Predicted \\
\hline
2011-10-16 & 55852.0 
        & 57.1  $\pm$ 0.7 & 57.3 $\pm$ 1.1 & 56.27 $\pm$ 0.03
        & 116.0 $\pm$ 0.9 & 35.2 $\pm$ 1.1 & 36.1 $\pm$ 0.1 \\
        
2011-10-17 & 55853.0 
        & 56.5  $\pm$ 1.2 & 56.6 $\pm$ 1.4 & 56.06 $\pm$ 0.03
        & 112.5 $\pm$ 1.4 & 31.7 $\pm$ 1.5 & 33.5 $\pm$ 0.1 \\
        
2011-10-19 & 55855.0 
        & 56.0 $\pm$ 0.4 & 56.1 $\pm$ 0.9 & 55.37 $\pm$ 0.03 
        & 108.2  $\pm$ 0.7 & 27.4 $\pm$ 1.0 & 28.3 $\pm$ 0.1 \\
\hline
2012-07-29 & 56139.0 
        & 39.2 $\pm$ 1.1 & 40.9 $\pm$ 2.2 & 41.86 $\pm$ 0.04 
        & 179.6 $\pm$ 4.7 & 113.5 $\pm$ 4.3 & 117.7 $\pm$ 0.2 \\
\hline
2012-12-19 & 56281.8 
		& 45.1 $\pm$ 1.0 & 46.1 $\pm$ 2.4 & 47.35 $\pm$ 0.04 
        & 52.8 $\pm$ 1.2 & 353.1 $\pm$ 3.5 & 356.0 $\pm$ 0.1 \\
2012-12-20 & 56282.8 
		& 44.3 $\pm$ 1.2 & 45.4 $\pm$ 2.5 & 46.42 $\pm$ 0.04 
        & 49.2 $\pm$ 1.5 & 349.6 $\pm$ 3.6 & 352.2 $\pm$ 0.1 \\
\hline
\label{tab_Capella_val}
\end{tabular}
\end{table*}

\subsection{The case of Capella}
\label{sec_cape_case}

The calibrated closure phases of Capella are shown in Fig. \ref{fig_CP_Capella}. As is readily apparent, large deviations from the null closure phase are detected. Large phase changes of $\sim \pm \pi$ indeed occur when the observed object is centrosymmetric, indicating here that the flux ratio between the two components is close to unity, as is expected for the Capella binary system at visible wavelengths. The fact that the transitions are smooth indicates that the flux ratio is not exactly equal to one. Thus these observations have clearly resolved the binary system.

However, some of these phase transitions occur in the opposite direction than expected by the best fit model (see the second closure phase graph in the first row of Fig. \ref{fig_CP_Capella} for instance). This effect is observed in around 6\,\% of all closure phase measurements (84$\times$2 closure phases per date of observation). A possible explanation is that the measurements may be affected by an uncalibrated bias on the imaginary part of the bispectrum. In this case, a sharp transition from 0 to $\pm \pi$ (imaginary part initially equal to 0) can therefore be turned into a smoother transition from 0 to $\pi$ or $-\pi$, depending on the sign of the bias offset.

This is most likely the reason why the minimum of the reduced $\chi^2$ function is generally much larger than 1. It reaches up to $\sim40$ (for one spectral channel) in the worst cases. Therefore, prior to computing the likelihood functions, the statistical error bars are scaled such that the minimum of the reduced $\chi^2$ function per spectral channel is 1 (error bars in the Fig. \ref{fig_CP_Capella} are rescaled \rouge{accordingly}). 

Another consequence of these transitions in the wrong direction is that it sharpens the transition when averaging closure phase datasets. The mean value of two transitions in opposite direction will indeed result in values globally closer to $\pm$ 180\degr or 0\degr. This can potentially translate into flux ratios that are \rouge{biased} 
towards unity. As detailed in the next section, the eventuality cannot be completely discarded when analyzing the spectral flux ratio.

It can be noted though, that averaging over the closure phase has been performed using complex phasors in order to take into account that the phase is known modulo $2\pi$. This is necessary to compute the correct phase, especially in the case of phases around $\pm180\degr$, but that obviously does not prevent such bias effect. However, the final flux ratio estimates are assumed to be less affected by this bias, as they result from the average over two independent measurements provided by the two sets of simultaneously recombined fibers, and also from the average over the different dates of observations.

Since the source of this bias is still unknown \citep[photon bias only affects the real part of the bispectrum as established by][]{Wirnitzer1985}, this effect becomes obvious only once the best fit model has been determined. There is thus no objective criterion that could allow to discriminate these biased data \textit{a priori}. Nonetheless, closure phase error bars resulting from the average over all datasets corresponding to one date of observation are necessary larger when there is an uncertainty on the transition direction. \rouge{As can be seen} in Fig.~\ref{fig_CP_Capella}, error bars are much larger around phase transitions. As a consequence, these points are less weighted in the fit.

\subsection{Orbit}
\label{seq_orbit}

The analysis of our results concerning the companion position requires a calibration of the ($u$,$v$) plane, that is a calibration of the baseline lengths and orientations \citep[as explained by][under the imaging baseline section]{Woillez2013}. For the fit, the baselines are conveniently defined by their theoretical lengths and orientation taken in a reference frame linked to the segmented mirror. However, to be scientifically useful, these need to be transformed into the angular separation of the binary and its position angle relative to North. Two aspects have to be considered for the calibration:
\begin{itemize} 
\item the magnification factor between the telescope pupil and the pupil that is imaged in the FIRST instrument can vary with the alignment ;
\item the rotation angle between the sky East-West / North-South orientation and the reference frame linked to the segmented mirror orientation.
\end{itemize}

The rotation of the pupil plane is due to the periscope sending the beam from the adaptive optics bench towards the hole through the FIRST bench. It uses a mirror combination working at angles different from 45\degr\ that therefore induces an unknown rotation of the pupil relative to the sub-pupils.

The calibration over all baselines can thus be performed by estimating two parameters: a scaling factor and a rotation angle. This is done by measuring the position of another well-known binary star, using the same data reduction pipeline as for Capella. Among the binary systems observed at the same epochs as Capella, the triple system of Algol ($\delta$ Per) is an ideal calibrator. The inner system (Algol A of type B8V and Algol B of type K2IV) of orbital period 2.87\,days and semi-major axis of 2.3\,mas is unresolved by a 3-m telescope. On the other hand, the outer system (Algol A+B and Algol C of type F1V) of orbital period \rouge{$\approx$} 
680\,days, semi-major axis of 93.8\,mas and flux ratio estimated to 10 at visible wavelengths is well detected by FIRST-18. Its orbital period is much longer than the Capella period (104\,days) and makes it a suitable target for the baseline calibration.

The relative positions of Algol A+B and C have therefore been estimated for several dates during the various runs, as shown in Table \ref{tab_Obslog}. The results are shown in Table \ref{tab_Algol}. The position parameters have been converted into polar coordinates in order to derive the angle for the rotation to be applied to the field of view. The predicted positions are computed from the orbit resulting from the measurements with the CHARA interferometer fitted by \cite{Baron2012}.

The Algol system was observed twice during both the October 2011 and December 2012 observing runs. In both cases, multiple observations of the systems within a given observing run yield consistent calibration factors at the 2$\sigma$ level or better (see Table \ref{tab_Algol}). This confirms the expectation that the rotation of the pupil in comparison with the FIRST sub-pupils is stable over the duration of an observing run since the injection part of the set-up was not modified during the course of each run. In October 2011, uncertainties on the calibration factors are much larger on October 16 than on October 19, probably as a consequence of changes in $r_0$ (see Table \ref{tab_Obslog}) and short atmospheric coherence time. While the latter could not be measured, the shorter integration times used on October 16, which results from a compromise between sensitivity and fringe visibility, is indicative of a shorter timescale for atmospheric turbulence. We therefore use the estimated position of Algol C regarding Algol A+B on October 19 to correct for the pupil rotation and scaling factor for all three dates of observation in October 2011. For the December 2012 observing run, we use the weighted average of both observations of Algol since they have similar precision. 

\begin{figure}
\centering
\resizebox{\hsize}{!}{\includegraphics{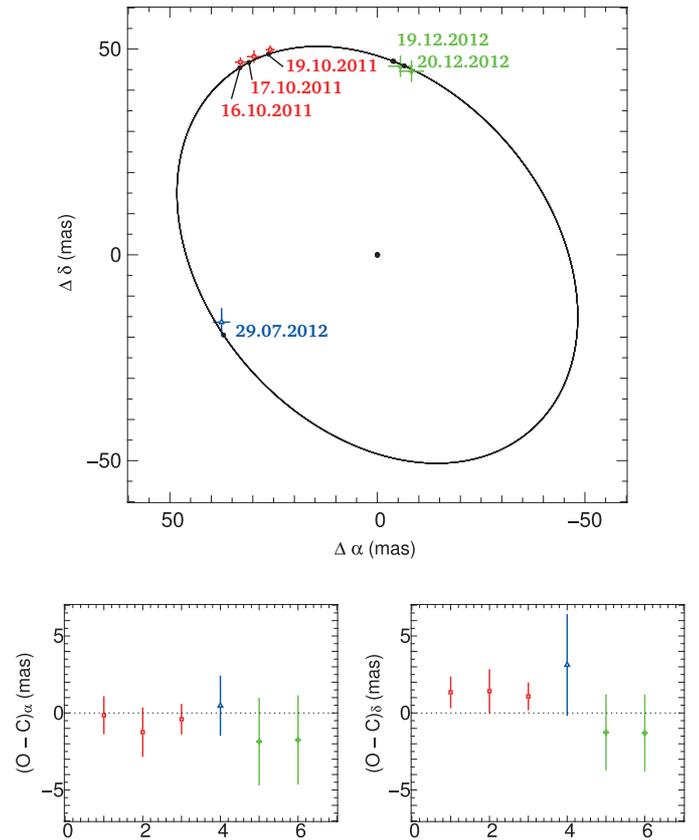}}
\caption{Capella companion positions relative to the predicted orbit. Top: Measured positions at different epochs: red for October 2011, blue for July 2012 and green for December 2012. Expected companion positions are marked as black dots and measurement points are drawn with error bars. Bottom: Difference between the estimated and expected values of the projection of the binary separation along the right ascension and declination axes. Each point corresponds to one observation date ranged in chronological order: 2011 October 16, 17 and 19, 2012 July 29 and 2012 December 19 and 20.}
\label{fig_Capella_orbit}
\end{figure}

The final relative astrometry of the Capella system throughout our observations is shown in Fig. \ref{fig_Capella_orbit}. The orbit computed from the model parameters fitted by \cite{Torres2009} is drawn for comparison. Table \ref{tab_Capella_val} presents both uncorrected and corrected (field rotation and scaling factor taken into account) estimates, along with their associated uncertainties. Most estimates are in agreement with the predicted positions within the 1-$\sigma$ range. \rouge{However}, the positions measured in 2011 seem to be affected by a small systematic error of about 2\,mas along the N-S direction (see Fig. \ref{fig_Capella_orbit}, bottom). It is plausible that this results from the baseline calibration, since all three points have been calibrated by the same astrometric estimate for the Algol system. Ideally, the calibrator should have been observed with sufficient accuracy at the three different dates.

\subsection{Spectral model}

\begin{figure*}
\sidecaption
\includegraphics[width=12cm]{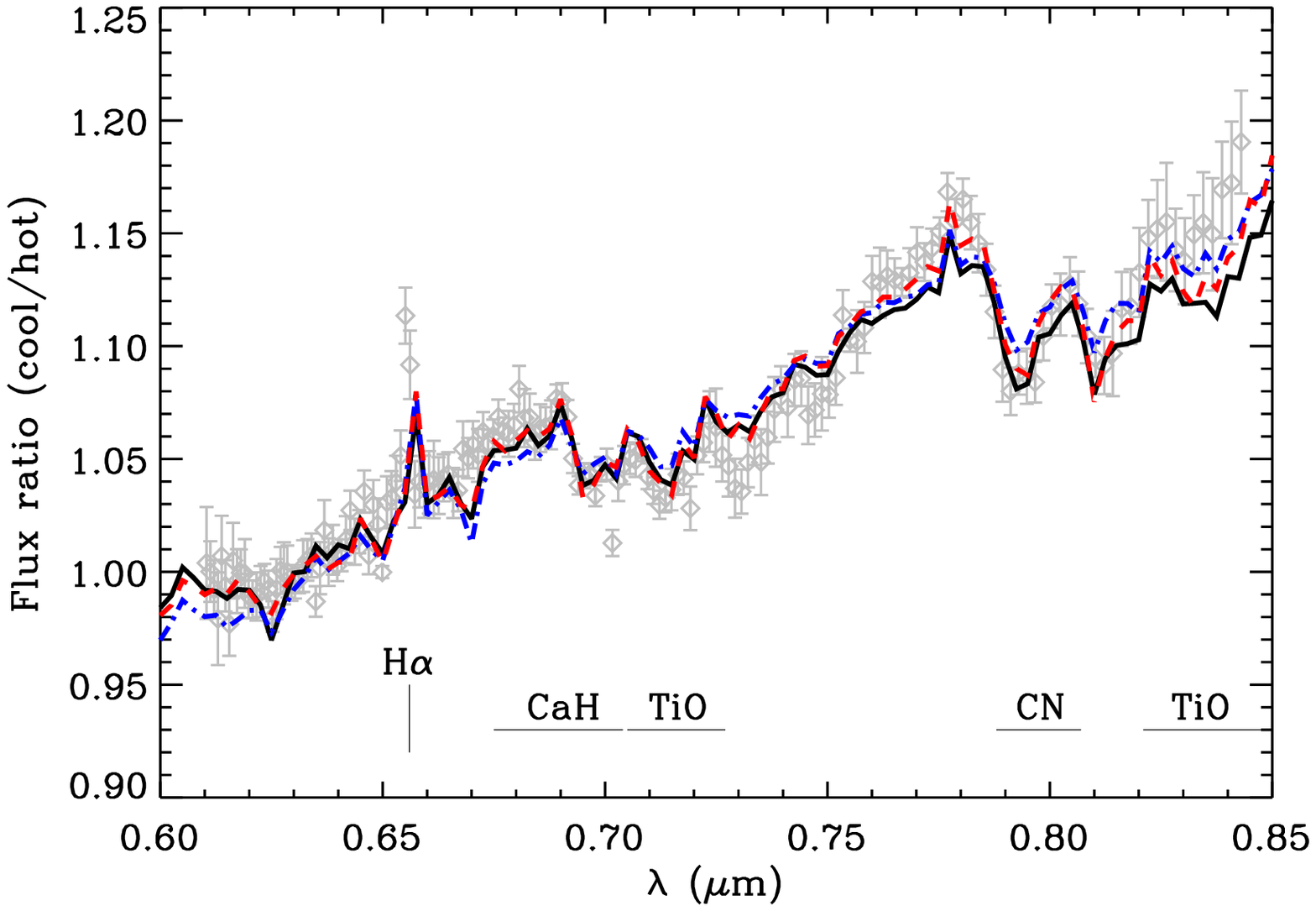}
\caption{Flux ratio for the Capella binary as a function of wavelength as measured with FIRST (gray diamonds). All curves represent predicted flux ratio spectra based on the \texttt{PHOENIX} grid of models and the set of solar abundances from \cite{Asplund2009}. The blue dot-dashed curve shows the predicted flux ratio spectrum for the stellar parameters inferred by \cite{Torres2009}, while the solid black curve represents the formal best fit (in the $\chi^2$ sense) to the flux ratio spectrum from the solar metallicity model grid. The red dashed curve shows the best fit for the model grid for a metallicity of $[m/{\rm H}] = +0.5$. Stellar parameters for all three models are listed in Table \ref{tab_stell_param}. The main atomic and molecular features that drive the model fitting are identified.}
\label{fig_Capella}
\end{figure*}

Along with the astrometry of the binary, our analysis also yields a "spectrum" of the binary flux ratio, $\rho(\lambda)$. We have inspected the quality of the flux ratio spectra from each individual night and each of the two separate V-grooves. As expected given seeing conditions, the flux ratios measured on December 2012 are of much worse quality than those from October 2011. Similarly, the July 2012 flux ratio spectrum is very noisy, owing to observations taken during early-morning twilight. While the astrometric position estimates from these epochs benefit from the average over all spectral channels and are thus reasonably accurate, the flux ratio spectra obtained during the 2012 runs have been discarded for the spectral analysis. We therefore averaged all October 2011 datasets to produce a final flux ratio  spectrum for the Capella binary system. The resulting final spectral resolution is about 1.5\,pixel as determined from data taken using a monochromatic laser beam. Uncertainties on the flux ratio spectrum were estimated as the standard deviation of the mean over the datasets. These results are available at the \textit{CDS} (column~1 lists the wavelengths in nm, column~2 and 3 give the flux ratio and their uncertainty\rouge{, }respectively).

The flux ratio spectrum obtained with FIRST agrees with historical measurements of the binary taken within our wavelength range at the $\lesssim 2 \sigma$ level. Furthermore, our spectrally dispersed observations reveal an overall slope that is consistent with data over a broader range and confirm that the cool component of the system is increasingly brighter at longer visible wavelengths. We find that the flux ratio reversal occurs at a wavelength of 0.64$\pm$0.01\,$\mu$m. In addition to its general slope, the FIRST flux ratio spectrum reveals finer structures, such as a sharp peak at 0.655\,$\mu$m and two broad dips around 0.69--0.75\,$\mu$m and 0.78--0.84\,$\mu$m. Both can be traced to the difference in effective temperature between the two stars\rouge{, }which leads to photospheric features of different depths: the former is H$\alpha$ (the slight mismatch in wavelength results from a mild inaccuracy in our wavelength solution) while the latter two are molecular bands (most notably TiO and CN) that are characteristic of cool photospheres. These spectral features are a direct confirmation that the infrared-bright component is cooler than the visible-bright one.

To quantify the constraint on stellar properties provided by the FIRST flux ratio spectrum, we use the most recent BT-Settl models\footnote{http://phoenix.ens-lyon.fr/fallard/} partially published in a review by \citep{Allard2012a} and described by \cite{Allard2012b}. These model atmospheres are computed with the \texttt{PHOENIX} multi-purpose atmosphere code version 15.5 \cite{Allard2001} solving the radiative transfer in 1D spherical symmetry, with the classical assumptions: hydrostatic equilibrium, convection using the mixing length theory, chemical equilibrium, and a sampling treatment of the opacities. The models use a mixing length as derived by the radiation hydrodynamic simulations of \cite{Ludwig2002,Ludwig2006} and \cite{Freytag2010,Freytag2012} and a radius as determined by the \cite{Baraffe1998} interior models as a function of the atmospheric parameters ($T_{\rm eff}$, $\log g$, $[m/{\rm H}]$). The reference solar elemental abundances used in this version of the BT-Settl models are those defined by \cite{Asplund2009}. For solar metallicity and higher, no \rouge{$\alpha$} element enhancement is required.

The \texttt{PHOENIX} library includes stellar spectra with a 100\,K and 0.5 sampling in $T_{\rm eff}$ and $\log g$, respectively. We explored the parameter space as follows: each model consisted on two pairs of stellar parameters ($T_{\rm eff}$ and $\log g$ for each component). The model flux ratio is computed using the same sampling and resolution as the FIRST data and normalized so as to match the observed median flux ratio. Since the \texttt{PHOENIX} emission spectra are given per unit surface area, this normalization is equivalent to setting the ratio of stellar radii to best match the median flux ratio in our spectral bandpass. In principle, our analysis could consider this ratio as a free parameter in the fit instead. However, since the FIRST flux ratios are in good agreement with previous flux ratio estimates of the binary, we do not expect to gain additional insight on the ratio of stellar radii. We therefore decided to remove one free parameter from our fit by using this \textit{a priori} normalization to ensure that the model fitting was primarily attempting at reproducing the global slope and finer spectral features, which depend only on the \rouge{stellar} effective temperatures and surface gravities. We believe that our data are more apt at constraining the latter. Nonetheless, we note that our normalization for the best fit models is consistent at the 1$\sigma$ level with the stellar radii ratio of $0.737\pm0.044$ estimated by \cite{Torres2009}.

We thus create a grid of flux ratio spectra that depends on four parameters, the effective temperature and surface gravity of both components while approximately maintaining the average stellar luminosity ratio in the FIRST bandpass. By performing a wide grid search around the stellar parameters derived by \cite{Torres2009}, the best fitting model 
has the following parameters: $T_{\rm eff}^{\rm hot}=5300$\,K, $T_{\rm eff}^{\rm cool}=4700$\,K, $\log g^{\rm hot} = \log g^{\rm cool} = 2.0$. This combination of stellar parameters yields a reasonably good match to the observed flux ratio spectrum (reduced $\chi^2 = 2.89$, see Fig. \ref{fig_Capella} and Table\,\ref{tab_stell_param}). We note that there is another peak appearing in the likelihood distribution, corresponding to a model with $T_{\rm eff}^{\rm hot}=5500$\,K, $T_{\rm eff}^{\rm cool}=4800$\,K and leading to a marginally poorer $\chi^2$, \rouge{which} provides a sense for the uncertainties on the derived stellar effective temperatures.

The stellar parameters derived here are significantly different from those estimated by \cite{Torres2009}, although the effective temperatures are only 4--7\% smaller than their nominal values. As shown in Fig. \ref{fig_Capella}, the higher effective temperatures proposed by \cite{Torres2009} also result in detectable molecular features in the flux ratio spectrum, although they are less marked than in our data (this is particularly true for the 0.78--0.84\,$\mu$m feature). This issue can be partially alleviated by using synthetic stellar spectra corresponding to lower surface gravity strengths. However, the overall quality of the fit is much poorer (see Table\,\ref{tab_stell_param}).

One solution to improve the fit consists in using a metal-rich composition. Using the series of \texttt{PHOENIX} models computed for $[m/{\rm H}] = +0.5$ (where elemental abundances are increased uniformly from the set of solar abundances), we find that the FIRST data are best fit using effective temperatures of 5600 and 4900\,K and surface gravities of $\log g^{\rm hot} = 2.5$ and $\log g^{\rm hot} = 2.5$. This model is significantly better (reduced $\chi^2 = 2.31$, see Fig. \ref{fig_Capella}) than the best fit model at solar metallicity, and has stellar properties in good agreement with \cite{Torres2009}. However, the super-solar metallicity is at odds with all estimates for the components of the system. We thus conclude that the observed flux ratio spectrum of the Capella binary does not conform perfectly to the prediction of stellar atmosphere models.

Breaking down the information provided by the FIRST flux ratio spectrum, it appears that the overall slope across the 0.6--0.85\,$\mu$m is reasonably well fit with both the nominal effective temperatures and the lower values preferred by the model fitting above, so long as the difference in effective temperatures between the two components is about 10--12\%. Thus it appears that the spectral features, specifically the depth of the molecular bands, represent the main culprit in forcing the fit away from the nominal stellar parameters. As mentioned in the previous section, we cannot exclude with certainty that a subtle bias in our data analysis is affecting the resulting flux ratio spectrum, possibly because the system is close to a unity flux ratio. However, we find it extremely unlikely that this bias could affect the wavelengths where molecular features are present but not the adjacent continuum. Indeed, such biases are expected to be present where the closure phases undergo large shift (from $\pm \pi$ to 0 for instance), as explained in the Section \ref{sec_cape_case}. This is therefore \rouge{expected to be} 
independent of spectral features, as the occurrence of these shifts only depends on the spatial frequencies ($\propto B/\lambda$, $B$ corresponding to the baseline) involved in the closure triangle. We thus believe that the mismatch between observations and models is an astrophysical effect instead.

Cooler effective temperatures, lower surface gravities and/or higher metallicity are all factors that result in deeper molecular bands in each component and in an increased difference in their strength, resulting in deeper features in the flux ratio spectrum as well. Since the stellar parameters of the Capella binary have been precisely determined, our FIRST flux ratio spectrum thus demonstrates that the molecular absorption bands in the spectra of each component are deeper than predicted by the models. One possible explanation for this shortcoming of the models is that the molecular opacities are underestimated in the models, because of incomplete line lists being used and/or underestimated oscillator strengths. Departures from local thermal equilibrium in stellar atmospheres can induce substantial effects for A-type stars but are negligible in the range of effective temperatures relevant to the Capella system. The two molecules that account for most of the opacity in the spectral features observed with FIRST are TiO and CN. For the former molecule, the version of the \texttt{PHOENIX} models used here are based on the line lists and oscillator strengths from \cite{Plez1998}. The CN line list and oscillator strengths are adopted from the SCAN database \citep{Jorgensen1990}. We believe that the TiO molecule, whose features are frequent in a broad range of cool stars, is much better calibrated and thus conclude that the reason for the under-prediction of these features by current atmospheric features most likely stems from the treatment of the CN molecule, either through the incompleteness of its line list or as a result of oscillator strengths that are too low.

\begin{table}
\centering
\caption{Stellar parameters of the \texttt{PHOENIX} atmospheric models used in this analysis from \cite{Allard2012b} and assuming the set of solar abundances from \cite{Asplund2009}. The first two models are at solar metallicity while the third one is assuming super-solar metallicity. All three models are shown in Fig.\,\ref{fig_Capella}. The parameters adopted to mimic those from \cite{Torres2009} are slightly different from those given in their paper because of the discrete sampling of the \texttt{PHOENIX} model grid. The last line of the table gives the ratio of stellar radii derived from the scaling factor used to normalize the spectra as a first step of the modelling.}
\begin{tabular}{cccc}
\hline\hline
& \multicolumn{2}{c}{$[m/{\rm H}] = 0$} & $[m/{\rm H}] = +0.5$\\
Parameter & nominal$^1$ & best fit & best fit \\
\hline
$T_{\rm eff}^{\rm hot}$  & 5700  & 5300  & 5600 \\
$T_{\rm eff}^{\rm cool}$ & 4900  & 4700  & 4900 \\
$\log g^{\rm hot}$       & 3.0   & 2.0   & 2.5 \\
$\log g^{\rm cool}$      & 2.5   & 2.0   & 2.5 \\
$\chi_{\rm red}^2$       & 4.37  & 2.89  & 2.31 \\
\hline
$R^{\rm cool}/R^{\rm hot}$ & 0.703 & 0.723 & 0.726 \\
\end{tabular}
\tablebib{$^1$ \citet{Torres2009}.}
\label{tab_stell_param}
\end{table}

\section{Summary of findings and conclusion}

In order to demonstrate the capabilities of a fibered aperture masking instrument like FIRST to provide valuable spectrally-dispersed information on binary systems whose separation is on the order of the diffraction limit, the results presented in this paper are focused on the binary star Capella. Its separation is indeed \rouge{comparable to} 
the diffraction limit and its flux ratio close to unity at visible wavelengths. Capella has been observed at three different epochs between 2011 and 2012 with FIRST-18 mounted on the 3-m Shane telescope of Lick Observatory (using its adaptive optics system as a fringe tracker). The secondary component has been detected at, or slightly below, the diffraction limit of the telescope at visible wavelengths with an accuracy well below a tenth of the diffraction limit. This first achievement \rouge{illustrates} 
the high angular resolution capability of the instrument. 

Using FIRST, we have also {\it directly} measured, for the first time, the flux ratio of the binary system at a spectral resolution of $R\sim300$ between 600 and 850\,nm. This spectral range gives access to spectral features (H$\alpha$ line, TiO and CN bands) that are quite influential when comparing the observed flux ratio spectrum with predictions based on \texttt{PHOENIX} library of synthetic spectra. The effective temperatures derived from this analysis are slightly offset (by 5--7\%) from those estimated by \cite{Torres2009} based on the extensive literature on this system. While we cannot exclude a subtle bias \rouge{affecting} 
our flux ratio measurements arising from the fact that the flux ratio is close to unity, this discrepancy probably \rouge{indicates} 
that the photospheric models used to predict the synthetic spectra are based on incomplete line lists and/or underestimated oscillator strengths for molecules commonly found in G- and K-type giants (most likely CN). This conclusion illustrates the power of FIRST in bringing valuable spectral information to characterize binary systems.

\begin{acknowledgements}

The authors would like to thank the staff from the Lick Observatory who provided an efficient and friendly support, especially in the effort of mounting the FIRST instrument and during the observing nights: Keith Baker, Bob Owen, Erik Kovacs, Kostas Chloros, Donnie Redel, Wayne Earthman, Paul Lynam and Pavl Zachary. They are also grateful to Dr. Bolte, Director of the  University of California Observatories, for his commitment to the project and generous telescope time allocation. They also thank the students from UC Berkeley who helped during the observing runs: S. Goeble and K. J. Burns, or helped improve the data reduction software: B. Bordwell. Dr. Helmbrecht, President and Founder of Iris AO, is also greatly thanked for his precious support concerning the segmented mirror. E. Huby would like to thank Alain Delboulbé for his valuable experience and recommendation concerning the optical bench dedicated to equalize the fiber lengths. Finally, we acknowledge financial support from Programme National de Physique Stellaire (PNPS) of CNRS/INSU, France and from a Small Research Grant of the American Astronomical Society. F. Marchis contribution to this work was supported by NASA Grant NNX11AD62G and by the National Science Foundation under Award Number AAG-0807468.
\end{acknowledgements}

\bibliographystyle{aa}
\bibliography{FIRST_biblio}

\begin{thebibliography}{39}
\expandafter\ifx\csname natexlab\endcsname\relax\def\natexlab#1{#1}\fi

\bibitem[{{Allard} {et~al.}(2001){Allard}, {Hauschildt}, {Alexander},
  {Tamanai}, \& {Schweitzer}}]{Allard2001}
{Allard}, F., {Hauschildt}, P.~H., {Alexander}, D.~R., {Tamanai}, A., \&
  {Schweitzer}, A. 2001, \apj, 556, 357

\bibitem[{{Allard} {et~al.}(2012{\natexlab{a}}){Allard}, {Homeier}, \&
  {Freytag}}]{Allard2012a}
{Allard}, F., {Homeier}, D., \& {Freytag}, B. 2012{\natexlab{a}}, Royal Society
  of London Philosophical Transactions Series A, 370, 2765

\bibitem[{{Allard} {et~al.}(2012{\natexlab{b}}){Allard}, {Homeier}, {Freytag},
  \& {Sharp}}]{Allard2012b}
{Allard}, F., {Homeier}, D., {Freytag}, B., \& {Sharp}, C.~M.
  2012{\natexlab{b}}, in EAS Publications Series, Vol.~57, EAS Publications
  Series, ed. C.~{Reyl{\'e}}, C.~{Charbonnel}, \& M.~{Schultheis}, 3--43

\bibitem[{{Anderson}(1920)}]{Anderson1920}
{Anderson}, J.~A. 1920, \apj, 51, 263

\bibitem[{{Asplund} {et~al.}(2009){Asplund}, {Grevesse}, {Sauval}, \&
  {Scott}}]{Asplund2009}
{Asplund}, M., {Grevesse}, N., {Sauval}, A.~J., \& {Scott}, P. 2009, \araa, 47,
  481

\bibitem[{Baldwin \& Haniff(2002)}]{Baldwin2002}
Baldwin, J.~E. \& Haniff, C.~A. 2002, Phil. Trans. R. Soc. London, Ser. A, 360
  no.1794, 969

\bibitem[{{Baraffe} {et~al.}(1998){Baraffe}, {Chabrier}, {Allard}, \&
  {Hauschildt}}]{Baraffe1998}
{Baraffe}, I., {Chabrier}, G., {Allard}, F., \& {Hauschildt}, P.~H. 1998, \aap,
  337, 403

\bibitem[{{Barlow} {et~al.}(1993){Barlow}, {Fekel}, \& {Scarfe}}]{Barlow1993}
{Barlow}, D.~J., {Fekel}, F.~C., \& {Scarfe}, C.~D. 1993, \pasp, 105, 476

\bibitem[{{Baron} {et~al.}(2012){Baron}, {Monnier}, {Pedretti}, {Zhao},
  {Schaefer}, {Parks}, {Che}, {Thureau}, {ten Brummelaar}, {McAlister},
  {Ridgway}, {Farrington}, {Sturmann}, {Sturmann}, \& {Turner}}]{Baron2012}
{Baron}, F., {Monnier}, J.~D., {Pedretti}, E., {et~al.} 2012, \apj, 752, 20

\bibitem[{{Blasius} {et~al.}(2012){Blasius}, {Monnier}, {Tuthill}, {Danchi}, \&
  {Anderson}}]{Blasius2012}
{Blasius}, T.~D., {Monnier}, J.~D., {Tuthill}, P.~G., {Danchi}, W.~C., \&
  {Anderson}, M. 2012, \mnras, 426, 2652

\bibitem[{{Campbell}(1899)}]{Campbell1899}
{Campbell}, W.~W. 1899, \apj, 10, 177

\bibitem[{{Chang} \& {Buscher}(1998)}]{Chang1998}
{Chang}, M.~P. \& {Buscher}, D.~F. 1998, in Society of Photo-Optical
  Instrumentation Engineers (SPIE) Conference Series, ed. R.~D. {Reasenberg},
  Vol. 3350, 2--13

\bibitem[{{Cieza} {et~al.}(2013){Cieza}, {Lacour}, {Schreiber}, {Casassus},
  {Jord{\'a}n}, {Mathews}, {C{\'a}novas}, {M{\'e}nard}, {Kraus}, {P{\'e}rez},
  {Tuthill}, \& {Ireland}}]{Cieza2013}
{Cieza}, L.~A., {Lacour}, S., {Schreiber}, M.~R., {et~al.} 2013, \apjl, 762,
  L12

\bibitem[{{Evans} {et~al.}(2012){Evans}, {Ireland}, {Kraus}, {Martinache},
  {Stewart}, {Tuthill}, {Lacour}, {Carpenter}, \& {Hillenbrand}}]{Evans2012}
{Evans}, T.~M., {Ireland}, M.~J., {Kraus}, A.~L., {et~al.} 2012, \apj, 744, 120

\bibitem[{{Freytag} {et~al.}(2010){Freytag}, {Allard}, {Ludwig}, {Homeier}, \&
  {Steffen}}]{Freytag2010}
{Freytag}, B., {Allard}, F., {Ludwig}, H.-G., {Homeier}, D., \& {Steffen}, M.
  2010, \aap, 513, A19

\bibitem[{{Freytag} {et~al.}(2012){Freytag}, {Steffen}, {Ludwig},
  {Wedemeyer-B{\"o}hm}, {Schaffenberger}, \& {Steiner}}]{Freytag2012}
{Freytag}, B., {Steffen}, M., {Ludwig}, H.-G., {et~al.} 2012, Journal of
  Computational Physics, 231, 919

\bibitem[{{Grady} {et~al.}(2013){Grady}, {Muto}, {Hashimoto}, {Fukagawa},
  {Currie}, {Biller}, {Thalmann}, {Sitko}, {Russell}, {Wisniewski}, {Dong},
  {Kwon}, {Sai}, {Hornbeck}, {Schneider}, {Hines}, {Moro Mart{\'{\i}}n},
  {Feldt}, {Henning}, {Pott}, {Bonnefoy}, {Bouwman}, {Lacour}, {Mueller},
  {Juh{\'a}sz}, {Crida}, {Chauvin}, {Andrews}, {Wilner}, {Kraus}, {Dahm},
  {Robitaille}, {Jang-Condell}, {Abe}, {Akiyama}, {Brandner}, {Brandt},
  {Carson}, {Egner}, {Follette}, {Goto}, {Guyon}, {Hayano}, {Hayashi},
  {Hayashi}, {Hodapp}, {Ishii}, {Iye}, {Janson}, {Kandori}, {Knapp}, {Kudo},
  {Kusakabe}, {Kuzuhara}, {Mayama}, {McElwain}, {Matsuo}, {Miyama}, {Morino},
  {Nishimura}, {Pyo}, {Serabyn}, {Suto}, {Suzuki}, {Takami}, {Takato},
  {Terada}, {Tomono}, {Turner}, {Watanabe}, {Yamada}, {Takami}, {Usuda}, \&
  {Tamura}}]{Grady2013}
{Grady}, C.~A., {Muto}, T., {Hashimoto}, J., {et~al.} 2013, \apj, 762, 48

\bibitem[{Haniff {et~al.}(1987)Haniff, Mackay, Titterington, Sivia, Baldwin, \&
  Warner}]{Haniff1987}
Haniff, C.~A., Mackay, C.~D., Titterington, D.~J., {et~al.} 1987, Nature, 328,
  694

\bibitem[{Hanuschik(2003)}]{Hanuschik2003}
Hanuschik, R.~W. 2003, \aap, 407, 1157

\bibitem[{Hinkley {et~al.}(2011)Hinkley, Carpenter, Ireland, \&
  Kraus}]{Hinkley2011}
Hinkley, S., Carpenter, J.~M., Ireland, M., \& Kraus, A.~L. 2011, The
  Astrophysical Journal, 730, L21

\bibitem[{Huby {et~al.}(2012)Huby, Perrin, Marchis, Lacour, Duch{\^e}ne,
  Choquet, Gates, Woillez, Lai, F{\'e}dou, Collin, Chapron, Arslanyan, \&
  Burns}]{Huby2012}
Huby, E., Perrin, G., Marchis, F., {et~al.} 2012, A\&A, 541, A55

\bibitem[{{Hummel} {et~al.}(1994){Hummel}, {Armstrong}, {Quirrenbach},
  {Buscher}, {Mozurkewich}, {Elias}, \& {Wilson}}]{Hummel1994}
{Hummel}, C.~A., {Armstrong}, J.~T., {Quirrenbach}, A., {et~al.} 1994, \aj,
  107, 1859

\bibitem[{{Jorgensen} \& {Larsson}(1990)}]{Jorgensen1990}
{Jorgensen}, U.~G. \& {Larsson}, M. 1990, \aap, 238, 424

\bibitem[{Lacour {et~al.}(2007)Lacour, Thi\'ebaut, \& Perrin}]{Lacour2007}
Lacour, S., Thi\'ebaut, E., \& Perrin, G. 2007, MNRAS, 374, 832

\bibitem[{Lacour {et~al.}(2011)Lacour, Tuthill, Amico, Ireland, Ehrenreich,
  Huelamo, \& Lagrange}]{Lacour2011}
Lacour, S., Tuthill, P., Amico, P., {et~al.} 2011, A\&A, 532

\bibitem[{{Lagrange} {et~al.}(2012){Lagrange}, {Milli}, {Boccaletti}, {Lacour},
  {Thebault}, {Chauvin}, {Mouillet}, {Augereau}, {Bonnefoy}, {Ehrenreich}, \&
  {Kral}}]{Lagrange2012}
{Lagrange}, A.-M., {Milli}, J., {Boccaletti}, A., {et~al.} 2012, \aap, 546, A38

\bibitem[{Le~Bouquin \& Absil(2012)}]{Lebouquin2012}
Le~Bouquin, J.-B. \& Absil, O. 2012, A\&A, 541, A89

\bibitem[{{Ludwig} {et~al.}(2002){Ludwig}, {Allard}, \&
  {Hauschildt}}]{Ludwig2002}
{Ludwig}, H.-G., {Allard}, F., \& {Hauschildt}, P.~H. 2002, \aap, 395, 99

\bibitem[{{Ludwig} {et~al.}(2006){Ludwig}, {Allard}, \&
  {Hauschildt}}]{Ludwig2006}
{Ludwig}, H.-G., {Allard}, F., \& {Hauschildt}, P.~H. 2006, \aap, 459, 599

\bibitem[{{Merrill}(1922)}]{Merrill1922}
{Merrill}, P.~W. 1922, \apj, 56, 40

\bibitem[{{Millour} {et~al.}(2004){Millour}, {Tatulli}, {Chelli}, {Duvert},
  {Zins}, {Acke}, \& {Malbet}}]{Millour2004}
{Millour}, F., {Tatulli}, E., {Chelli}, A.~E., {et~al.} 2004, in Society of
  Photo-Optical Instrumentation Engineers (SPIE) Conference Series, ed. W.~A.
  {Traub}, Vol. 5491, 1222

\bibitem[{{Newall}(1899)}]{Newall1899}
{Newall}, H.~F. 1899, \mnras, 60, 2

\bibitem[{Perrin {et~al.}(2006)Perrin, Lacour, Woillez, \&
  Thi\'ebaut}]{Perrin2006}
Perrin, G., Lacour, S., Woillez, J., \& Thi\'ebaut, E. 2006, MNRAS, 373, 747

\bibitem[{{Plez}(1998)}]{Plez1998}
{Plez}, B. 1998, \aap, 337, 495

\bibitem[{{Sanchez-Bermudez} {et~al.}(2012){Sanchez-Bermudez}, {Sch{\"o}del},
  {Alberdi}, \& {Pott}}]{Sanchez-Bermudez2012}
{Sanchez-Bermudez}, J., {Sch{\"o}del}, R., {Alberdi}, A., \& {Pott}, J.~U.
  2012, Journal of Physics Conference Series, 372, 012025

\bibitem[{Torres {et~al.}(2009)Torres, Claret, \& Young}]{Torres2009}
Torres, G., Claret, A., \& Young, P.~A. 2009, Astrophysical Journal, 700, 1349

\bibitem[{{Weber} \& {Strassmeier}(2011)}]{Weber2011}
{Weber}, M. \& {Strassmeier}, K.~G. 2011, \aap, 531, A89

\bibitem[{Wirnitzer(1985)}]{Wirnitzer1985}
Wirnitzer, B. 1985, Journal of the Optical Society of America A, 2, 14

\bibitem[{{Woillez} \& {Lacour}(2013)}]{Woillez2013}
{Woillez}, J. \& {Lacour}, S. 2013, \apj, 764, 109

\end{thebibliography}

\end{document}